\documentclass[12pt]{article}
\usepackage{amsthm}
\usepackage{amsmath}
\usepackage{amsfonts}
\usepackage{amssymb}
\usepackage{amscd}
\usepackage{enumerate}
\usepackage[automark]{scrpage2}          

\def\>{\rightarrow}
\def\-{\mapsto}
\def\I{\mathbb{I}}
\def\N{\mathbb{N}}

\def\R{\mathbb{R}}
\def\C{\mathbb{C}}
\def\H{\mathbb{H}}
\def\P{\mathbb{P}}
\def\a{\alpha}
\def\b{\beta}
\def\c{\gamma}
\def\l{\lambda}
\def\p{\psi}

\def\2{\frac{1}{2}}
\def\3{\frac{1}{3}}
\def\4{\frac{1}{4}}
\def\8{\frac{1}{8}}

\def\be{\begin{equation}}
\def\ee{\end{equation}}
\def\bp{\begin{proof}}
\def\ep{\end{proof}}
\def\bc{\begin{cases}}
\def\ec{\end{cases}}

\newcommand{\bra}[1]{\ensuremath{\left\langle #1\right|}}
\newcommand{\ket}[1]{\ensuremath{\left|#1\right\rangle}}
\newcommand{\braket}[2]{\ensuremath{\left\langle #1\vphantom{#2}\right.\left|\vphantom{#1}#2\right\rangle}}
\newcommand{\tbraket}[2]{\ensuremath{\left\langle #1\vphantom{#2}\right.\left.\otimes\,\,\vphantom{#1}#2\right\rangle}}
\def\expect#1{\ensuremath{\langle{#1}\rangle}} 

\begin{document}


\title{Classical tensors from quantum states}

\author{P. Aniello$^{1,}$$^2$, G. Marmo$^1$, G. F. Volkert$^1$\\
{\emph {\small $^1$Dipartimento di Scienze Fisiche dell'Universit\`a di Napoli Federico II}}\\
{\emph {\small and Istituto Nazionale di Fisica Nucleare (INFN) -- Sezione di Napoli}}\\
{\emph {\small Complesso Universitario di Monte S. Angelo, via Cintia, I-80126 Napoli, Italy.}}\\
{\emph {\small $^2$Facolt\`a di Scienze Biotecnologiche, Universit\`a di Napoli Federico II}}\\
{\small E-mail: aniello@na.infn.it, marmo@na.infn.it, volkert@na.infn.it}}

\maketitle

\begin{abstract}
The embedding of a manifold $M$ into a Hilbert-space $\mathbb{H}$ induces, via the pull-back, a tensor field on $M$ out of the Hermitian tensor on $\mathbb{H}$.  We propose a general procedure to compute these tensors in particular for manifolds admitting a Lie-group structure.
\end{abstract}

\section{Introduction}
The geometrical identification of mathematical structures of quantum mechanics goes back to Dirac \cite{Dirac:1936}
\cite{Dirac:1945}, with the introduction of quantum Poisson brackets, and to Weyl, Segal and Mackey \cite{Weyl:1927vd}
\cite{Weyl:1931vd}
\cite{Segal:1947}
\cite{Mackey:1962}
\cite{Mackey:1957}
\cite{Mackey:1988}
\cite{Mackey:1998} who identified the role of the symplectic structure both in quantum mechanics and quantum field theory. In the same circle of ideas one may include the paper by Strocchi \cite{Strocchi:1956}. A strict geometrical formulation, however, is more recent (Heslot, Rowe, Cantoni, Cirelli et al., Ashtekar, Gibbons, Brody, Hughston, de Gosson) 
\cite{Heslot:1985}
\cite{Rowe:1980}
\cite{Cantoni:1975}
\cite{Cantoni:1977}
\cite{Cantoni:1977}
\cite{Cantoni:1980} 
\cite{Cantoni:19}
\cite{Cirelli:1983} 
\cite{Cirelli:1984}
\cite{Abbati:1984}
\cite{Ashtekar:1997ud}
\cite{Gibbson:1992}
\cite{Brody:2001}
\cite{De Gosson:2007},
and it has also been used systematically to introduce and analyze in the quantum setting the role of bi-Hamiltonian description of evolution equations
\cite{Carinena:2000}  
\cite{Marmo:2002}.\\
The geometrical formulation of quantum mechanics in the Dirac approach goes along the following lines. As a first step, one  replaces the Hilbert space $\mathbb{H}$ with the tangent bundle $T\mathbb{H}$ constructed on the real differential Hilbert manifold $\H^{\R}:=\mbox{Re}(\mathbb{H})\oplus\mbox{Im}(\mathbb{H})$, which replaces the usual complex separable Hilbert space. The Hermitian inner product $\mathbb{H}\times \mathbb{H}\rightarrow \mathbb{C}$ on quantum states is then replaced by an Hermitian tensor on quantum-state-valued sections of the tangent bundle $T\mathbb{H}$ defining a Riemannian tensor for the real part, and a symplectic structure for the imaginary part. Roughly speaking, this amounts to identify the Hilbert space $\mathbb{H}$ with the tangent space $T_{\phi}\mathbb{H}$ at each point $\phi$ of the base manifold. By using the Hermitian strucure one may start equally well with $\H^*$, the dual vector space of $\H$. By using sections of $T^*\mathbb{H}$ we define a Riemannian tensor in contravariant form and a Poisson tensor (i.e. a symplectic form written in contravariant form). \\
It has been remarked several times that in this formulation, quantum evolution described by the Schr\"odinger equation defines a Hamiltonian vector field which, in addition, preserves a complex structure and a related Riemanian metrics. With other words, vector fields representing quantum systems are not only Hamiltonian, they are also Killing vector fields. These remarks point out that several aspects of Hamiltomian dynamics may be also used with advantages in connection with quantum mechanics. On the other hand, it is well known that classical dynamics is fully described by using symplectic manifolds or, more generally, by means of Poisson manifolds when constraints are also taken into account. However on a selected family of quantum states parametrized by a real differential manifold $M$ some shadows of the additional structures existing in quantum mechanics appear also in this "classical framework" (This point of view has been emphasized many times by J. Klauder \cite{Klauder:1994}\cite{Klauder:1997}\cite{Klauder:2001}\cite{Klauder:2005}).\\ 
In this paper, we would like to investigate how to define classical tensors from quantum states by considering in particular manifolds admitting a Lie-Group structure $G \cong M$. Thus our procedure may be considered as way to implement Klauder's point of view.  This approach is closely related to the mathematical setting appearing in the generalized coherent states \cite{Perelomov:1971}\cite{Gilmore:1972}\cite{Onofri:1974}.  The identified manifold will depend on the initial fiducial state we start with to define the orbit. In this way we will have the possibility to define embeddings $M \hookrightarrow \mathbb{H}$ and the associated pull-backs in a natural framework by using unitary (vector or ray) representations $G \rightarrow U(\mathbb{H})$.  To clearly illustrate our procedure, we shall consider first the action of a group $G$ on a finite dimensional Hilbert-space. Then we shall move to the more realistic case of infinite dimensions by means of specific examples rather than dealing with general aspects. Depending on the particular nature of these Lie-group-manifolds we may identify them with configuration spaces or phase spaces. Of course, these constructions are not without relations with the quantum-classical transition, nevertheless they should not be considered equivalent to the classical counterpart. To be specific, finite-level quantum systems prevalently considered for quantum computing and quantum information are systems defined on finite dimensional manifolds, or even stratified real differential manifolds \cite{Grabowski:2005}, and do not correspond to any classical limit - for this reason one should avoid considering our procedure as a way to explicitly define classical dynamical systems corresponding to quantum ones. Moreover the tensor fields defined here will have only a
kinematical interpretation. The family of quantum evolutions or transformations admitting a counterpart on the finite dimensional manifolds has to be analyzed separately. 

\section{Hermitian tensor fields on the Hilbert manifold}	
We consider a separable complex Hilbert space $\H$. On its realification $\H^{\R}$ we construct then tangent and cotangent bundles $T\H^{\R}$ and $T^{*}\H^{\R}$. To introduce coordinate functions we use an orthonormal basis $\{\ket{e_j}\}_{j\in \mathcal{J}}$ where the index set may be finite or infinite dimensional ($\mathcal{J}:=\N$). For any vector $\ket{\p}$ we set:
\be  z^j(\p) = \braket{e^j}{\p} = q^j(\p)+ip^j(\p)\,.\ee 
Usually we shall simply write $z^j$ or $(q^j, p^j)$, respectively, for complex or real coordinates, and drop the argument. When we need to use a continuous basis, say for the coordinate or momentum representation, we write \be \ket{\p}= \int dx \ket{x}\braket{x}{\p}=\int dx \ket{x}\p(x)\ee
with $dx$ representing the Lebesque-measure. In what follows our statements should be considered to be always mathematically well defined whenever the Hilbert space we are considering is finite dimensional. In the case of an infinite-dimensional Hilbert space, additional qualifications are needed whenever we have to deal with unbounded operators; these cases will be handled separately when it is the case instead of making general claims. 
To make computations easy to follow we shall use symbols like $d\ket{\p} := \ket{d\p}$. They should be understood as defining vector-valued differential forms, i.e. 
\be \ket{d\p} = d(z^j\ket{e_j}) = dz^j \ket{e_j}\,.\ee
Specifically, we assume that an orthonormal basis has been selected once and it does not depend on the base point. To deal with "moving frames" one should introduce a connection. Of course this can be done and in  some specific situations it is useful and convenient. This is the case when we deal with Berry phases or the non-abeliean generalization of Wilczek and Zee \cite{Berry:1984} \cite{Wilczek:1984}. \\
In this respect, $\ket{d\p}$ should be thought of as a section of the cotangent bundle $\H \rightarrow T^{*}\H \cong \H \times \H^{*}$ tensored with the Hilbert space $\H$. With this notation, the usual Hermitian inner product
\be \braket{\p}{\p}= \braket{z^j e_j}{z^k e_k} = \braket{e_j}{e_k} \bar{z}^j z^k = \delta_{jk}\bar{z}^j z^k \ee  
is easily promoted to a tensor field (Hermitian or K\"ahlerian tensor field) by setting 
\be\tbraket{d\p}{d\p}:= \braket{e_j }{e_k } d\bar{z}^j\otimes dz^k=\delta_{jk}d\bar{z}^j\otimes dz^k\,.\label{HT}
\ee
By factoring out real and imaginary part we find:
\be\tbraket{d\p}{d\p} = \delta_{jk}(dq^j\otimes dq^k+dp^j\otimes dp^k)+i\delta_{jk}(dq^j\otimes dp^k-dp^j\otimes dq^k)\,.\label{Re+Im HT}
\ee
Thus the Hermitian tensor decomposes into an Euclidean metric and a symplectic form. Clearly, infinitesimal generators of one-parameter groups of unitary transformations will be at the same time Hamiltonian vector fields. In this sense for quantum evolution we may be able to use most of the mathematical tools which have been elaborated for Hamiltonian dynamics (Arnold, Abraham-Mardsen, Marmo et al., Lieberman-Marle) \cite{Arnold:76}\cite{Abraham:78}\cite{marmo:85}\cite{Libermann:87}.\\
If with any vector $\ket{\p}$ we associate a vector field  
\be X_{\p}: \H \>  T\H;\,\phi \- (\phi, \p),\ee
then it is possible to write a contravariant Hermitian tensor which we may write in the form    
\be \tbraket{\frac{\delta}{\delta\p}}{\frac{\delta}{\delta\p}}:= \braket{e_j}{e_k}
\frac{\partial}{\partial \bar{z}^j}\otimes\frac{\partial}{\partial z^k}\ee
Again, the decomposition into real and imaginary part would give
\be 
\delta^{jk}(\frac{\partial}{\partial q^j}\otimes\frac{\partial}{\partial q^k}+\frac{\partial}{\partial p^j}\otimes\frac{\partial}{\partial p^k} )\ee 
for the real part, and
\be \delta^{jk}(\frac{\partial}{\partial q^j}\otimes\frac{\partial}{\partial p^k} -\frac{\partial}{\partial p^j}\otimes\frac{\partial}{\partial q^k} )\,,\ee
for the imaginary part. As the probabilistic interpretation of quantum mechanics requires that the identification of quantum states is made of with rays of $\H$ (one dimensional complex vector spaces) rather than with vectors, our tensors should be defined on the ray space $\mathcal{R}(\H)$; i.e. the complex projective space $\C\P (\H)$, instead of $\H$.
Equivalence classes of vectors are defined by $\p \sim \varphi$ iff $\p = \lambda \varphi$ for $\lambda \in \C_0:= \C-\{0\}$. At the infinitesimal level, the action of the group $\C_0$ is generated by the vector fields
\be
\triangle := q^j\frac{\partial}{\partial q^j}+ p^j\frac{\partial}{\partial p^j}: \H \>  T\H;\,\p \- (\p, \p)
\ee
and 
\be
\Gamma := p^j\frac{\partial}{\partial q^j}+ q^j\frac{\partial}{\partial p^j}:\H \>  T\H;\,\p \- (\p, J\p)\,.
\ee
Here $J$ is the one-one tensor field representing the complex structure on the realified version of the complex Hilbert space.
For contravariant tensor fields $\tau$ on $\H$ to be projectable onto $\mathcal{R}(\H)$ one has to require that $L_{\triangle}\tau= 0 $ and $L_{\Gamma}\tau= 0 $. On the other hand, for covariant tensor fields, to be the pull-back of tensor fields on $\mathcal{R}(\H)$ it is necessary that $L_{\triangle}\alpha =0,\, L_{\Gamma}\alpha =0$ and moreover $i_{\triangle}\alpha =0, \, i_{\Gamma}\alpha =0$.\\
These remarks allow to conclude that the Hermitian tensor on $\H_{0}$ (the Hilbert space $\H$ without the zero vector), which is the pull-back of the K\"ahlerian tensor on $\mathcal{R}(\H)$, has the form 
\be
\frac{\tbraket{d\psi}{d\psi}}{\braket{\psi}{\psi}}- \frac{\braket{\psi}{d\psi}}{\braket{\psi}{\psi}}\otimes \frac{\braket{d\psi}{\psi}}{\braket{\psi}{\psi}}\,.\label{PHT}\ee
For further details on these tensors we refer the reader to 
\cite{Cirelli:1983} 
\cite{Cirelli:1984}
\cite{Abbati:1984}
\cite{Ashtekar:1997ud}
\cite{Carinena:2000}
\cite{Clemente:2008}.

\section{Tensors on Lie groups from finite dimensional representations}
Let us consider a Lie group $G$ acting on a vector space $V$. This means that there exists a Lie-Group homomorphism
\be \pi: G  \> \mbox{Aut}(V) \ee 
or an action 
\be \phi: G \times V \> V\,. \ee 
For each fiducial vector $v_0 \in V$ we define a submanifold in $V$ given by
\be \phi(G\times \{v_0\})= \{\pi(g)\cdot v_0\}\subset V\,. \ee 
By considering the tangent bundle construction, we find an action of $TG$ on $TV = V\times V$
\be T\phi: TG \times TV \> TV\,. \ee
Because $TG \cong G \times \mathfrak{g}$, $\mathfrak{g}$ being the Lie-Algebra of $G$, the tangent map $T\phi$ requires the existence of a representation on $V$ of the Lie algebra $\mathfrak{g}$.
In general, in the finite dimensional case, the representation of $\mathfrak{g}$ extends naturally to a representation of the enveloping algebra $\mathcal{U}(\mathfrak{g})$ on $V$. In infinite dimensions when we start with unitary representations of $G$ on $\H$, the fiducial vector should be chosen to be smooth or analytic, so that again we have a natural extension of the representation of the Lie-algebra to the enveloping algebra \cite{Nelson:1959}.\\
The main idea we use to construct covariant tensors on $G$ out of covariant tensors on $V$ is to consider the map
\be \phi_{v_0}: G \> V\ee  
as an embedding so that we may pull-back to $G$ the algebra of functions $\phi_{v_0}^{*}(\mathcal{F}(V))\subset \mathcal{F}(G)$, and, along with the relation connecting with the exterior differential on the two spaces
\be d\phi_{v_0}^{*}=\phi_{v_0}^{*}d\,,\ee  
we are able to pull-back all the algebra of exterior forms, but also the tensor algebra generated by one-forms with real or complex valued functions as coefficients.\\
When the group acts directly on the space of rays, $\mathcal{R}(V)$,  for a complex vector space $V$, by means of 
\be\phi(g)\cdot [v] = [\pi(g)\cdot v]\,,\ee 
the corresponding action on $V$, by means of $\pi(g)$, need not be a true representation but it is enough that is defined up to a multiplier, i.e.   
\be\pi(g)\cdot\pi(h)= m(g,h)\pi(g,h)\ee 
with $m(g,h)$ a non zero complex number. Thus the quantum mechanical probabilistic interpretation does not require that $\pi$ is a vector representation but only that it is a ray-representation. In many cases we have to deal with this additional freedom.\\
In his seminal paper \cite{Bargmann}, Bargmann associated a vector-representation of a central extension of $G$ by means of the multiplier $m$ (the so called Bargmann group of $G$) with a ray representation of $G$. The most important example is provided by the Abelian vector group which may be centrally extended to the Heisenberg-Weyl group. Another important example is provided by the Galilei group.\\
Let us start with a vector representation of $G$ on a vector space $V$. The orbit of the action of $G$ on $V$, starting with the fiducial vector $v_0$ will be denoted by $M=\phi(G\times \{v_0\}) \subset V$. We shall use for convenience the bra-ket notations of Dirac. We have 
\be U(g)\ket{0}=\ket{g};\,\,\{\ket{g}\}_{g\in G}=M\,. \ee       
It should be noticed that $M$ will not be a vector space and may be given a manifold structure by using the differential structure on $G$. If $G_0$ is the isotropy group of $\ket{0}$, we find $M:= G/G_0$. The vectors parametrized by $M$ may generate the full vector space by means of linear combinations. We may use an orthonormal basis for $V$ and define coordinate functions $z^j(g) = \braket{e_j}{g}$,
The vector-valued one-forms we obtain by taking the exterior derivative
\be d \ket{g} = d U(g)\ket{0} = dU(g) U^{-1}(g)\ket{g}\ee
and the Hermitian tensor $\tbraket{d\p}{d\p}$, when calculated on the manifold $M$ (the pulled-back tensor) will be 
\be  \tbraket{dg}{dg}:=\bra{g}(dU(g) U^{-1}(g))^{\dagger}\otimes dU(g) U^{-1}(g)\ket{g}\,. \ee \label{HTG}
If we denote by $X^1, X^2,..., X^n$ the generators of the left action of $G$ on itself, i.e. the right invariant infinitesimal generators and by $\theta_1, \theta_2,...\theta_n$ the corresponding dual basis of one forms, i.e. $\theta_j(X^k)=\delta_j^k$, we consider $U(t)=e^{itR(X)}$ and we find
\be dU(g)U^{-1}(g)= iR(X^j)\theta_j \ee
along with
\be (dU(g)U^{-1}(g))^{\dagger}= -iR(X^j)\theta_j \ee
because the infinitesimal generators are skew-Hermitian. In conclusion:
\be \tbraket{dg}{dg} = \bra{g}R(X^j)R(X^k)\ket{g}\theta_j\otimes\theta_k\,.\ee
By decomposing the basis elements $\theta_j\otimes\theta_k$ into
\be \2(\theta_j\otimes \theta_k+\theta_j\otimes \theta_k)+\2(\theta_j\otimes \theta_k-\theta_j\otimes \theta_k):=\2\theta_j\odot \theta_k+\2\theta_j\wedge \theta_k\,,\ee 
it is possible to extract the real part 
\be \2\bra{g}R(X^j)R(X^k)+R(X^k)R(X^j)\ket{g}\theta_j\odot\theta_k \ee
and the imaginary part 
\be \2\bra{g}R(X^j)R(X^k)-R(X^k)R(X^j)\ket{g}\theta_j\wedge\theta_k \ee
in the usual way.
Because the commutator of Hermitian operators is skew hermitian, the second term is imaginary and we have derived a Riemannian tensor along with a (pre-)symplectic structure. It should be remarked that the Riemannian tensor is the expectation value of an element of order two in the homomorphic image, provided by the representation of the enveloping algebra of the Lie algebra $G$.
Thus in the infinite dimensional situation we have to consider whether $\ket{g}$ is in the domain of the operator of order two which appears in the definition of the Riemannian tensor. Some theorems are avaible \cite{Nelson:1959} \cite{Stinespring:1959} \cite{Davies:1971}, but we shall not be concerned with these problems here. 
By using the fact that $R$ is associated with the tangent map of a vector or ray unitary representation, we find 
\be R(X^j)R(X^k)-R(X^k)R(X^j)=iR([X^j,X^k])+i\omega(X^j,X^k),\ee
where $\omega$ is a closed 2-form on the group associated with the multiplier $m$ when we deal with a ray representation instead of a vector representation of $G$.
\\\\
\emph{Remark:} By using the adjoint action of $G$ on its Lie algebra, it is possible to go from right invariant vector fields to left-invariant ones. In this way the expectation values of operators generated by right-invariant infinitesimal generators on the states $\ket{g}$ may be replaced by the expectation values of the corresponding operators written in terms of left-invariant infinitesimal generators evaluated on the initial fiducial state $\ket{0}$.
\\\\
If we introduce $Y^1,Y^2,...Y^n$ generators of the right action along with\\$\a_1,\a_2,...,\a_n$ dual one-forms, $\a_j(Y^k)=\delta_j^k$, we have also 
\be \tbraket{dg}{dg} = \bra{0}R(Y^j)R(Y^k)\ket{0}\a_j\otimes\a_k\,.\ee 
In this way the role of the fiducial vector and the requirement that it should be in the domain of the operators of order two in the enveloping algebra of the left invariant generators becomes more clear. It may be convenient to derive in general form the pull-back K\"ahlerian tensor when we start with an action on the ray space, the complex projective space, instead of the Hilbert space.\\
Here we have to start not with $\tbraket{d\p}{d\p}$ in (\ref{HT}) but with   
\be
\frac{\tbraket{d\psi}{d\psi}}{\braket{\psi}{\psi}}- \frac{\braket{\psi}{d\psi}}{\braket{\psi}{\psi}}\otimes \frac{\braket{d\psi}{\psi}}{\braket{\psi}{\psi}}\,\ee
in (\ref{PHT}). Therefore the pulled-back tensor becomes
\be
\frac{\tbraket{dg}{dg}}{\braket{g}{g}}- \frac{\braket{g}{dg}}{\braket{g}{g}}\otimes \frac{\braket{dg}{g}}{\braket{g}{g}}\,\ee
After simple computations we find 
\be
\Bigg(\frac{\bra{g}R(Y^j)R(Y^k)\ket{g}}{\braket{g}{g}}- \frac{\bra{g}R(Y^j)\ket{g}}{\braket{g}{g}} \frac{\bra{g}R(Y^k)\ket{g}}{\braket{g}{g}}\Bigg)\theta_j\otimes\theta_k\,\ee
The net result is that the closed 2-form will not be effected, except for the normalization, while the metric tensor will be modified by the addition of an extra term
\be \bra{g}R(Y^j)\ket{g}\bra{g}R(Y^k)\ket{g}\theta_j\odot\theta_k\ee
Few comments are in order. From the expression of the $jk-$th coefficient of the pulled back tensor 
\be \bra{0}R(Y^j)R(Y^k)\ket{0}-\bra{0}R(Y^j)\ket{0} \bra{0}R(Y^k)\ket{0}\ee
we notice that when
\be R(Y^k)\ket{0} = \lambda^k\ket{0}\ee
we find 
\be \lambda^k\bra{0}R(Y^j)\ket{0} - \bra{0}R(Y^j)\ket{0}\lambda^k=0\,.\ee
It means that the subalgebra of $\mathfrak{g}$ of the subgroup of $G$ which acts on $\ket{0}$ simply by multiplication by a phase will give rise to "degeneracy directions" for the Hermitian tensor. In more specific terms the tensor we are pulling back provides a tensor on $G/G_0$, i.e. on the homogeneous space defined by the isotropy subgroup (up to a phase) of the fiducial vector.\\
In the coming sections we are going to consider some specific examples which have been selected because of their relevance for physical problems. 

\section{Pulled-back tensors on a compact space: $SU(2)$}
The simplest non-trivial compact Lie-group is given by $SU(2)\cong S^3$. An embedding of this group into the Hilbert space 
\be \H=\mathrm{L}^2(SU(2)):=\bigoplus_{s} \C^{2s+1}, \mbox{s integer or half integer}\ee 
can be realized in different ways, since it will depend on the choice of the spin-$s$-representations 
\be U^s: SU(2)\> \mbox{Aut}(\C^{2s+1}),\,\, g \mapsto U^s(g).\ee
By using
\be dU^s(g)^{\dagger}=-U^s(g)^{\dagger}dU^s(g)U^s(g)^{\dagger}\ee
the pulled back tensor reads
\be \bra{0}(dU^s(g))^{\dagger}\otimes dU^s(g)\ket{0}= \bra{0}R^s(Y^j)R^s(Y^k)\ket{0}\theta_j\otimes \theta_k\,.\ee
The pulled back tensor associated to the pulled back tensor on $SU(2)/U(1)\cong S^2$ provides on the other hand the structure
\be \big(\bra{0}R^s(Y^j)R^s(Y^k)\ket{0} - \bra{0}R^s(Y^j)\ket{0}\bra{0}R^s(Y^k)\ket{0}\big)\theta_j\otimes \theta_k\,.\ee
If we choose $\ket{0}$ to be an eigenvector of $R^s(Y^3)$ it follows that the symmetric tensor and skew-symmetric form are both degenerate.\\  
Let us compute this pulled back tensors in the defining representation $s=1/2$ with $R(Y^j):=\sigma^j$, the Pauli-matrices explicitly. Here we get based on the fiducial state 
\be \ket{0}:=\left(\begin{array}{c}1 \\0\end{array}\right) \in \C^2  \ee
a symmetric tensor
\be  \2\bra{0}\sigma^j\sigma^k+\sigma^k\sigma^j\ket{0}\theta_j\odot\theta_k=\2\delta^{jk}\theta_j\odot\theta_k= \2\theta_j\otimes\theta_j  \label{RSU(2)}\ee
and an antisymmetric tensor 
\be  \frac{1}{2i}\bra{0}\sigma^j\sigma^k-\sigma^k\sigma^j\ket{0}\theta_j\wedge\theta_k= - d\theta_3\,,\label{S-SU(2)} \ee
where we have used the decomposition 
\be \sigma^j\sigma^k = \delta^{jk}\sigma^0 + i\epsilon_r^{jk}\sigma^r\label{T-jk}\,,\ee 
and the Maurer-Cartan relation \be d\theta_r +\2{c_r}^{jk}\theta_j\wedge \theta_k= 0\ee
with ${c_r}^{jk}={\epsilon_r}^{jk}$ for $G=SU(2)$.
By using furthermore the right invariant one-forms on $SU(2)$ given by
\begin{equation}\theta_1= \sin(\alpha)d\beta - \sin(\beta)\cos(\alpha)d\gamma\,,\label{1-u2}\end{equation}
\begin{equation}\theta_2= \cos(\alpha)d\beta + \sin(\beta)\sin(\alpha)d\gamma\,,\label{2-u2}\end{equation}
\begin{equation}\theta_3= d\alpha + \cos(\beta)d\gamma\,,\label{3-u2}\end{equation}
we see that the symmetric tensor (\ref{RSU(2)}) coincides with the Riemannian tensor
\be \2 (d\alpha \otimes d\alpha +d\beta\otimes d\beta + 2 \cos(\beta)d\alpha\odot d\gamma),\label{Euclid on S3}\ee
which is induced on a three-sphere by an embedding in a four dimensional Euclidean space, where else its "projective" counterpart 
\be \big(\bra{0} \sigma^j\sigma^k\ket{0} - \bra{0}\sigma^j\ket{0}\bra{0}\sigma^k\ket{0}\big)\theta_j\otimes \theta_k\,,\ee
coincides after symmetrization with the induced metric   
\be \2 (d\beta\otimes d\beta + \sin^2(\beta)d\c\odot d\c)\,,\label{Euclid on S2}\ee
on a two-sphere being embedded in three dimensional Euclidean space. 
The antisymmetric part (\ref{S-SU(2)}) turns out to be equal to  
\be \sin(\beta)d\b\wedge d\c\,. \ee

\section{Weyl systems and pulled-back tensors on a symplectic vector space}
We consider now a symplectic vector space $(V, \omega)$. A Weyl system is defined by a map from $V$ to the set of unitary operators on a Hilbert space $\H$. This map is required to be strongly continuous and satisfying the following properties: 
\begin{enumerate}
\item $W(v)\in U(\H)\,,$ for all $v\in V$\,; 
\item $W(v_1)W(v_2)W^{\dagger}(v_1)W^{\dagger}(v_2)= e^{i\omega(v_1,v_2)}\I\,.$
\end{enumerate}
Here $\omega$ is the symplectic structure on $V$ \cite{esposito:04}. The symplectic structure is the "infinitesimal form" of the multiplier $m(v_1,v_2)$ appearing in the definition of "ray-representations" for the abeliean vector group $V$ \cite{Aniello:2000}. It should be remarked that for different orderings, the symplectic structure is actually replaced by an Hermitian product on $V$. Let us now carry on the general procedure on this specific example - For simplicity we introduce a basis in $V$, say $\{e_1,e_2, ..., e_{2n}\}$, so that $v=v^{j}e_j$.\\
With the help of the Stone-von Neumann theorem it is possible to write 
\be W(v)=e^{iR(v)}\,.\ee 
In particular this relation implies
\be [R(v_1),R(v_2)]= i\omega(v_1,v_2)\ee
Now, after the selection of a fiducial vector $\ket{0}$, we have to compute 
\be \bra{0}(dW)^{\dagger}\otimes dW\ket{0}\ee
First we notice that unitarity of $W(v)$ implies that $d(W^{\dagger})=(dW)^{\dagger}$. Than, by using the decomposition $v=v^je_j$, we find for the pulled back tensor:
\be \bra{0}R(e_j)R(e_k)\ket{0} dv^j\otimes dv^k\,.\ee
By considering the real part and the imaginary part respectively, we find
\be \2\bra{0}R(e_j)R(e_k)+R(e_k)R(e_j)\ket{0} dv^j\odot dv^k\label{Re-PB-HT on R2n}\ee
and
\be \frac{1}{2i}\bra{0}R(e_j)R(e_k)-R(e_k)R(e_j)\ket{0} dv^j\wedge dv^k= \omega_{jk} dv^j\wedge dv^k\,.\label{Im-PB-HT on R2n}\ee
If, as we should for physical interpretation, we consider the pull-back of the K\"ahlerian tensor from the complex projective space associated with $\H$, we should find the same imaginary term, but the symmetric part should be evaluated as
\be \2\bra{0}R(e_j)R(e_k)+R(e_k)R(e_j)\ket{0}-\bra{0}R(e_j)\ket{0}\bra{0}R(e_k)\ket{0}\,.\label{Re-PB-PHT on R2n}
\ee 
From this expression it follows clearly that the fiducial vector $\ket{0}$ should be selected such that it belongs to the domain of $R(e_j)$ for all $j\in \{1,2,...2n\}$ and to the domain of the elements of order two in the enveloping algebra.
\\\\
\emph{Remark:} Note that, when we consider only an Abelian vector subgroup of $V$  which is a Lagrangian subspace, the pull-back tensor only contains the Euclidean part because $\omega_{jk}$ restricted to the subgroup will vanish identically.
\\\\
To evaluate (\ref{Re-PB-HT on R2n}) and (\ref{Re-PB-PHT on R2n}) we have to give a realization of $\H$. This is done by considering the decomposition of the symplectic vector space into $V=\R^n\oplus (\R^n)^*$. In the realization $\H=\mathrm{L}^2(\R^n)$ we may compute the expectation values of $R(e_j), R(e_k)$ and combination of these based on a Gaussian function 
\be\ket{0}:= Ne^{-\frac{1}{2}q^2}\in \mathrm{L}^2\cap C^{\infty}(\R^n)\,.\ee
Here we get due to the realizations
\be R(e_j)\ket{0}:=Q^j\ket{0}= Q^j(Ne^{-\frac{1}{2}q^2}) = q_j \ket{0}\ee
for $j\in \{1,2,...,n\}$ and
\be R(e_j)\ket{0}:=P^j\ket{0}= i\frac{\partial}{\partial q_j}(Ne^{-\frac{1}{2}q^2}) =-i q_j \ket{0}\ee
for $j\in \{n+1,n+2,...,2n\}$ the $\mathrm{L}^2(\mathbb{R}^n)$ inner products
\be \bra{0}Q^{j} P^k\ket{0} =-i\bra{0}  q_{j} q_k\ket{0} \ee
\be  \bra{0}P^{j} Q^k\ket{0} =i\bra{0}  q_{j} q_k\ket{0} \ee
\be  \bra{0}Q^{j} Q^k\ket{0} =\bra{0}  q_{j} q_k\ket{0} \ee
\be \bra{0}P^{j} P^k\ket{0} =\bra{0}  q_{j} q_k\ket{0}\,, \ee
which can be made explicit by the the integrals
\be I_{jk}:= \bra{0}  q_{j} q_k\ket{0} =N^2\int_{-\infty}^{\infty}d^nq e^{-q^2}q_jq_k\,.\ee
They get zero for $j\neq k$ due to 
\be I_{jk} = N^2\bigg(\int_{-\infty}^{\infty}dq_i e^{-q_i^2}\bigg)^{n-2} \bigg(\int_{-\infty}^{\infty}dq_j e^{-q_j^2} q_j \bigg)^2 =0 \ee
and non-zero for $j = k$ due to 
\be I_{jj} = N^2\bigg(\int_{-\infty}^{\infty}dq_ie^{-q_i^2}\bigg)^{n-1} \int_{-\infty}^{\infty}dq_j e^{-q_j^2} q_j^2  =   \frac{1}{2}N^2\pi^{n/2}.\ee  
By setting $N^2\pi^{n/2}\equiv 1$ we can summarise this into
\be I_{jk}=\2\delta_{jk}\ee
and since we have furthermore $\expect{P^j}= \expect{Q^j} = 0$ we can conclude that both relatations in (\ref{Re-PB-HT on R2n}) and (\ref{Re-PB-PHT on R2n}) define each of them a metric tensor field
\be g_{jk}dv^j\odot dv^k = \2 \delta_{jk} dv^j\odot dv^k, \ee
giving rise to an Euclidean metric on $\R^{2n}$.

\section{The pull-back on a manifold without group structure}
We consider a family of Hamiltonian operators $H(\l)$ with $\l \in M$, a smooth manifold and the eigenvalue problem 
\be H(\l) \ket{\p_0(\l)}=E_0(\l) \ket{\p_0(\l)},\ee
where $E_0$ defines the lowest nonzero eigenvalue which is supposed to be non-degenerate. This association $\l \mapsto \ket{\p_0(\l)}$ defines an embedding of $M$ into $\mathcal{R}(\H)$, the ray space of $\H$. Using the Hermitian tensor (\ref{PHT}), we find 
\be \frac{\tbraket{d\psi_0(\l)}{d\psi_0(\l)}}{\braket{\psi_0(\l)}{\psi_0(\l)}}- \frac{\braket{\psi_0(\l)}{d\psi_0(\l)}}{\braket{\psi_0(\l)}{\psi_0(\l)}}\otimes \frac{\braket{d\psi_0(\l)}{\psi_0(\l)}}{\braket{\psi_0(\l)}{\psi_0(\l)}}\,.\label{PB-HT on M}\ee 
The external derivative $d$ is meant to act on functions on $M$. By using $d=d\l^{\mu}\otimes \frac{\partial}{\partial \l^{\mu}}$, we find 
\be h_{\mu\nu}=\braket{\partial_{\mu}\p_0}{\partial_{\nu}\p_0}-\braket{\p_0}{\partial_{\nu}\p_0}\braket{\partial_{\mu}\p_0}{\p_0}\ee 
 with the requirement $\braket{\p_0}{\p_0}=1$. Using more generally the spectrum of $H(\l)$, say
 \be H(\l)\ket{a;\l}= E_a(\l)\ket{a;\l},\ee
 we have 
 \be dH(\l)\ket{a;\l}= dE_a(\l)\ket{a;\l}+E_a(\l)d\ket{a;\l}- H d\ket{a;\l}\,.\ee
 Taking the scalar product with $\bra{b;\l}$ we obtain
 \be \bra{b;\l}dH(\l)\ket{a;\l}=(E_a-E_b)\bra{b;\l}d\ket{a;\l}\,,\ee
 i.e.
\be d\ket{a;\l}=\sum_{b\neq a} \frac{\ket{b;\l}\bra{b;\l}dH\ket{a,\l}}{E_a-E_b}\ee
Using this expression for $a=0$, we get
\be d\ket{\p(\l)_0}=\sum_{b\neq a} \frac{\ket{b;\l}\bra{b;\l}dH\ket{\p{\l}_0}}{E_0-E_b}\,,\ee
which allows to write the pull-back of the Hermitian tensor given by (\ref{PB-HT on M}). It should be mentioned that a particular interesting application to physical systems has been provided by Zanardi et al.\cite{Zanardi}.

\section{Conclusions and outlook}
We have seen that, out of any  unitary representation of a group on some Hilbert space,
 it is possible to identify a manifold by acting with the group on some fiducial  
state. It is also possible to identify submanifolds by other procedures. On each submanifold,
it is possible to consider the pullback of the Hermitian tensor and therefore
obtain classical tensors out of the quantum states. In some sense, this procedure may give rise to a kind of \,"dequantization". As a matter of fact,  
this is not connected to any dynamics resp. quantum classical  
transition and we have to accept that there are quantum \emph{and}  
classical-like structures in every quantum system, which should be  
considered as coexistent. In particular, when we consider
the immersion of a symplectic vector space by means of a Weyl system, we obtain
not only the original symplectic structure but also an Euclidean tensor and therefore
a complex structure. With the help of this structure it is possible to define
complex coordinates and a correspondence between them and creation/annihilation operators. 
We should stress that while the symplectic structure turns out to be independent of
the fiducial vector we start with, the Riemannian tensor does depend on it.
More likey it is this particular aspect that makes the symplectic structure
more fundamental than the metric structure in classical mechanics.
On the other hand, when we consider the imbedding of a Lagrangian subspace
in the Hilbert space identified by a Weyl system, we find no symplectic structure 
but we find a metric tensor, this available metric tensor, intrinsically built,
permits to define the velocity field associated with a wave function in Bohmian mechanics.
Thus this procedure will allow us to define a Bohmian vector field on any
manifold we may immerse in the Hilbert space. In a future paper we shall consider more closely this problem and provide a general setting for Bohmian vector fields.

\section*{Acknowledgments}
This work was financially supported by the German Academic Exchange Service (DAAD) and the National Institute of Nuclear Physics (INFN).


\begin{thebibliography}{0}

\bibitem{Dirac:1936}
P. A. M. Dirac, \emph{The Principles of Quantum Mechanics} (Clarendon Press, Oxford, 1936). 

\bibitem{Dirac:1945}
P. A. M. Dirac, On the Analogy Between Classical and Quantum Mechanics, \emph{Rev. 
Mod. Phys.} {\bf 17} (1945), 195--199. 

\bibitem{Weyl:1927vd}
H. Weyl, Quantenmechanik und Gruppentheorie, \emph{Zeitsch. Phys.} {\bf 46} (1927), pp. 1--46. 

\bibitem{Weyl:1931vd}
H. Weyl, \emph{The theory of groups and Quantum Mechanics}, (Dover, New York, 1931). 

\bibitem{Segal:1947}
E. Segal, Postulates for general Quantum Mechanics, \emph{Ann. Math. }{\bf 48} (1947),  930--948. 

\bibitem{Mackey:1962}
G. W. Mackey, \emph{The mathematical foundations of Quantum Mechanics}, (W. A. Benjamin, New York, 1963). 

\bibitem{Mackey:1957}
G. W. Mackey, Quantum Mechanics and Hilbert space, The American Mathematica Monthly {\bf 64} (1957), pp. 45--57.  

\bibitem{Mackey:1988}
G. W. Mackey, Weyl's program and modern physics, \emph{Differential geometric 
methods in theoretical physics}, (Kluwer Acad. Publ., Dordrecht, 1988).

\bibitem{Mackey:1998} 
G. W. Mackey, The Relationship Between Classical mechanics and Quantum Mechanics, \emph{Contem. Math.} {\bf 214} (1998),  91--109.   

\bibitem{Strocchi:1956}
F. Strocchi, Complex coordinates and Quantum Mechanics, \emph{Rev. Mod. Phys.} {\bf 38} (1956),  36--40. 

\bibitem{Heslot:1985}
A. Heslot, Quantum mechanics as a classical theory, \emph{Phys. Rev. D} {\bf 31} (1985), 
1341--1348.  

\bibitem{Rowe:1980}
D. J. Rowe, A. Ryman and G. Rosensteel, Many body quantum mechanics 
as a symplectic dynamical system, \emph{Phys Rev A} {\bf 22} (1980),  2362--2372.  

\bibitem{Cantoni:1975}
V. Cantoni, Generalized transition probability, Comm. \emph{Math. Phys.} {\bf 44} (1975), 125--128.  

\bibitem{Cantoni:1977}
V. Cantoni, The Riemannian structure on the space of quantum-like systems, \emph{Comm. Math. Phys.} {\bf 56} (1977),  189--193.  

\bibitem{Cantoni:1977}
V. Cantoni, Intrinsic geometry of the quantum-mechanical phase space, Hamiltonian systems and Correspondence Principle, \emph{Rend. Accad. Naz. Lincei} {\bf 62} (1977),  628--636.
 
\bibitem{Cantoni:1980} 
V. Cantoni, Geometric aspects of Quantum Systems, \emph{Rend. sem. Mat. Fis. Milano} {\bf 48} (1980), 35--42.

\bibitem{Cantoni:19}
V. Cantoni, Superposition of physical states: a metric viewpoint, \emph{Helv. Phys. Acta} {\bf 58} (1985. ),  956--968. 
 
\bibitem{Cirelli:1983} 
R. Cirelli, P. Lanzavecchia and A. Mani\`a, Normal pure states of the von 
Neumann algebra of bounded operator as K\"ahler manifold, \emph{J. Phys. A: 
Math. Gen.} {\bf 16} (1983),  3829--3835.  

\bibitem{Cirelli:1984}
R. Cirelli and P. Lanzavecchia, Hamiltonian vector fields in Quantum 
Mechanics, \emph{Nuovo Cimento B} {\bf 79} (1984),  271--283. 

\bibitem{Abbati:1984}
M. C. Abbati, R. Cirelli, P. Lanzavecchia and A. Mani\`a, Pure states of 
general quantum mechanical systems as K\"ahler bundle, \emph{Nuovo Cimento 
B} {\bf 83} (1984),  43--60.  

\bibitem{Ashtekar:1997ud}
A. Ashtekar and T. A. Shilling, Geometrical formulation of Quantum 
Mechanics, in On EinsteinÕs path, Ed. A. Harvey (Springer, Berlin, 1998) 

\bibitem{Gibbson:1992}
G. W. Gibbson, Typical states and density matrices, \emph{J. Geom. 
Phys.} {\bf 8} (1992), 147--162.  

\bibitem{Brody:2001}
D. Brody and L.P. Hughston, Geometric quantum mechanics, \emph{J. Geom. 
Phys.} {\bf 38} (2001),  19--53.  

\bibitem{De Gosson:2007}
M. de Gosson, \emph{The principles of Newtonian and quantum mechanics} (Imperial College Press, London, 2001).

\bibitem{Carinena:2000}  
J. F. Carinena, J. Grabowski and G. Marmo, Quantum Bi-Hamiltonian Systems, \emph{Int. J. Mod. Phys. A} {\bf 15} (2000),  4797--4810. 

\bibitem{Marmo:2002}
G. Marmo, G. Morandi, A. Simoni and F. Ventriglia, Alternative Structures and 
bi-Hamiltonian Systems, \emph{J. Phys. A: Math. Gen.} {\bf 35} (2002),  8393--8406. 

\bibitem{Klauder:1994}
J. R. Klauder, Quantization without quantization, \emph{Ann. of Phys.} {\bf 237} (1995),  147--160. 

\bibitem{Klauder:1997}
J. R. Klauder, Understanding Quantization, \emph{Found. Phys.} {\bf 27} (1997), 1467--1483. 

\bibitem{Klauder:2001}
J. R. Klauder, Phase space geometry in classical and quantum mechanics,
quant-ph/0112010.

\bibitem{Klauder:2005}
D. Abernethy, J. R. Klauder, The distance between classical and quantum systems, \emph{Found. Phys.} {\bf 35}  (2005),  881--895.

\bibitem{Perelomov:1971}
A. M. Perelomov, Coherent States for Arbitrary Lie Groups, \emph{Commun. math. Phys.} {\bf 26} (1972),  222--236.

\bibitem{Gilmore:1972}
F. T. Arecchi, E. Courtens, R. Gilmore, H. Thomas, Atomic Coherent States in Quantum Optics,
\emph{Phys. Rev. A} {\bf 6} (1972), 2211--2237.
  
\bibitem{Onofri:1974}
E. Onofri, A note on coherent state representations on Lie groups, \emph{Journ. Math. Phys..} {\bf 16}  (1975), 1087.

\bibitem{Grabowski:2005}
J. Grabowski, M. Kus, G. Marmo, Geometry of quantum systems: density states and entanglement,  
 \emph{J. of Phys. A: Math. Gen.}, {\bf 38} (2005),  10217--10244.

\bibitem{Berry:1984} 
M. V. Berry, Quantal phase factors accompanying adiabatic changes, 
\emph{Proc. R. Soc. A} {\bf 392} (1984),  45--57.

\bibitem{Wilczek:1984}
F. Wilczek and A. Zee,  Appearance of Gauge Structure in Simple Dynamical Systems,
\emph{Phys. Rev. Lett.} {\bf 52} (1984),  2111--2114.

\bibitem{Arnold:76}
V. I. Arnold, \textit{Les methodes mathematiques de la Mecanique Classique} (Mir, Moscow 
1976).

\bibitem{Abraham:78}
R. Abraham, J. E. Marsden, \textit{Foundations of Mechanics} (Benjamin, Reading, MA, 
1978). 

\bibitem{marmo:85}  
G. Marmo, E. J. Saletan, A. Simoni, B. Vitale:
\textit{Dynamical Systems -- A Differential Geometric Approach to Symmetry and Reduction} (John Wiley \& Sons, Chichester, 1985).

\bibitem{Libermann:87}
P. Libermann, C. M. Marle, \textit{Symplectic Geometry and Analytical Mechanics}
(Reidel, Dordrecht, 1987). 

\bibitem{Clemente:2008}
J. Clemente-Gallardo, G. Marmo, Basics of Quantum Mechanics, Geometrization and some Applications to Quantum Information, \emph{Int. J. Geom. Meth. Modern Physics} {\bf 5} (2008).

\bibitem{Nelson:1959}
E. Nelson, Analytic Vectors, \emph{Ann. of Math., 2nd ser.}, {\bf 70} (1959), 572--615.

\bibitem{Bargmann}
V. Bargmann, On unitary ray representations of continuous groups, \emph{Ann. of 
Math.} {\bf 59} (1954),  1--46.

\bibitem{Stinespring:1959}
E. Nelson, Stinespring, W. Forest, Representation of elliptic operators in an enveloping algebra, \emph{Amer. J. Math.} {\bf 81} (1959), 547--560.

\bibitem{Davies:1971}
E. B. Davies, Hilbert space representations of Lie algebras,  \emph{Comm. Math. Phys.} {\bf 23} (1971), 159--168.

\bibitem{Aniello:2000}
P. Aniello, V. ManÕko, G. Marmo, S. Solimeno, F. Zaccaria, \emph{J. Opt. B: Quantum Semiclass. Opt.} {\bf 2} (2000), 718--725.

\bibitem{esposito:04}
G. Esposito, G. Marmo, G. Sudarshan: \textit{From Classical to Quantum Mechanics} (Cambridge University Press, Cambridge, 2004).

\bibitem{Zanardi}
P. Zanardi, P. Giorda, M. Cozzini, The differential information-geometry of quantum phase transitions,
quant-ph/0701061.

\end{thebibliography}
\end{document}